# Reality and locality in quantum mechanics, an experiment proposed

Jan Mycielski
Dept. of Mathematics
Univ. of Colorado
BOULDER, CO 80309-0395
E-mail address: Jan.Mycielski@Colorado.edu

**1. Introduction.** As well known, the standard (Copenhagen) interpretation of quantum mechanics (QM) presents real difficulties to its students. *First*, it suggests that in some circumstances, if the path of a particle was not observed, say, as a particle track, then this path is not real. *Second*, some instantaneous interactions between distant measurements, also called collapse of wave packets, defy the idea of time-space separation. These difficulties were clearly described e.g. by R. P. Feynman in [3] and [4]. In [4] he described intuitively path integration, and wrote that there are no waves but only particles, and even (on page 23) that "the wave theory collapsed". But, A. Zee pointed out in his introduction to [4], that Feynman's gave a new way of calculating wave functions, i.e., probability amplitudes, but he did not eliminate them.

The purpose of the present paper is to propose an experiment that could confirm in a direct way that either both difficulties are inescapable or to prove that, at least in certain areas of QM, the Copenhagen interpretation is wrong and a theory proposed by Max Planck (see T. S. Kuhn [5], Chapter X, and references therein) explains correctly a realm of quantum phenomena. A vast majority of physicists is convinced that the latter is out of the question, that Planck's model is false. They accept that certain phenomena are absolutely unpredictable and locality fails, just *as if we were observing miracles performed by an agency that exists outside of physical space-time*. Planck's theory claims otherwise, at least for some chapters of QM.

Several demonstrations of the difficulties mentioned above, similar to the one proposed in this paper, were made and all were consistent with the Copenhagen interpretation; see e.g. J. F. Clauser et al [2] and later A. Aspect et al [1]. But they were more complicated and hence less convincing than the experiment that will be propose here.

Concerning the history of this issue, as much as I know, in spite of the fact that in their time Planck's view was not clearly refuted, neither Einstein and Schrödinger who were skeptical about the Copenhagen interpretation nor Bohr who defended it, ever discussed Planck's theory in their writings. Perhaps they thought that it was not general enough to be useful. Nowadays the idea is hardly ever mentioned; the monograph of Kuhn [5] describes how it faded, but it does not explain fully why that happened. Thus I think that it would be valuable to confirm or to refute Planck in the direct way proposed below.

**2. Philosophical preparation.** From a methodological point of view the standard form of QM and Planck's idea are very different. The first is only a system of algorithms and the second is a theory or model. It is worthwhile to explain this difference since in the scientific literature the terms *theory*, *model*, and *algorithm* are often used in ambiguous ways; it is only in mathematical logic that these terms are precisely defined.

(1) QM *is not a model (and does not define a model) in any natural sense.*

*Argument corroborating* (1). QM talks about two kinds of objects, particles and waves, that influence each other but it does not represent them together in a single dynamical system.

(2) QM *is not a scientific theory in the sense of mathematical logic.*

This thesis (2) follows immediately from (1) plus a more general thesis.

(3) *Every scientific theory yields a model that satisfies all its theorems, a model that is natural in a set-theoretic sense.*

We will show that (3) follows immediately from two theses (4) and (5) to be stated momentarily. Let ***K*** denote the class of set theories *S* formalized in first-order logic that appear to be reasonable extensions of the standard first-order set theory ZFC. For example, extensions by some large cardinal axioms; see several references gathered in [6].

(4) *It is routine to interpret any scientific theory in a sufficiently strong set theory S of the class **K**.* (As a rule definitional extensions of ZFC suffice.)

*Argument corroborating* (4). The thesis (4) is supported by the vast experience of mathematics and of all sciences. The evidence grew from 1908 till 1923 such that (4) became convincing and remains fully convincing till today. Roughly speaking the main steps were the following. In 1908 Zermelo formulated the axioms for Cantor's set theory. Then Skolem gave a first-order formalization of Zermelo's axioms. Finally, in 1923, Hilbert published his definitional extension of first-order logic by means of $\varepsilon$-symbols. In consequence it became routine to translate informal mathematical definitions and proofs into the $\varepsilon$-extension of the first-order set theoretic language. Since that time several large-cardinal axioms analogous to the axiom of infinity of ZFC were invented. These axioms strengthen ZFC and they support beautiful new mathematics (see [6]).

(5) *For every set theory S of class **K** it is acceptable (whenever needed) to strengthen S by adding an axiom A(S) which tells "There exists an ordinal number $\alpha$ such that S is true in the natural model $\langle V_\alpha, \in \rangle$", and the theory S + A(S) belongs to **K**.* (For a definition of the natural models $\langle V_\alpha, \in \rangle$ see e.g. the penultimate paper listed in [6].)

*Argument corroborating* (5). The thesis (5) holds since it derives from the intended meaning of *S*.

(Recall that, if *S* is consistent, *S* + *A*(*S*) is a proper extension of *S*. Indeed *S* + *A*(*S*) implies that *S* is consistent, hence, by Gödel's Second Incompleteness Theorem, *A*(*S*) is not provable in *S*.)

*Argument corroborating* (3). As already mentioned, (3) follows immediately from (4) and (5).

Thus we have shown (1) that QM is not a model and (2) that it is not a scientific theory in the sense of mathematical logic. However, as we shall see in the next section, the proposal of Planck is a theory (however its scope may be limited), and it explains a number of experimental facts in quantum mechanics. The purpose of this paper is to define a real experiment that should refute or confirm Planck's theory in a straightforward way.

*Terminology.* As we have seen above, to have an interpretation of a theory $T$ in some $S$ in **K** is the same as to have a definition of a model $M$ of $T$ in some $S'$ in **K**. Thus I will use interchangeably the terms *model* and *theory*.

*Additional Remarks*. 1. I have not seen in the literature an explicit statement of the thesis (4). But I think that it is important to stress (4) since it refutes some ideas about mathematics expressed by popular writers; e.g. I. Lakatos, P. J. Davis and R. Hersh.

2. In mathematics every conjecture that has not been interpreted and proved in some $S$ of class **K** is an open problem. Likewise, the consistency of every theory, whose interpretation in any such $S$ is problematic, is an open problem. Quine's system NF is an example of such an unsettled theory.

3. By Gödel's Second Incompleteness Theorem, interpretability in some S of class **K** is the only possible *practical* test for the consistency of theories.

4. We observe that almost all mathematicians of all races and cultures who are familiar with the axioms of ZFC think that these axioms are *obvious* properties of the universe of sets that they imagine, and they apply these axioms *instinctively* in their proofs. This psychological fact demonstrates that ZFC with $\varepsilon$-symbols (ZFC + $\varepsilon$), is only a slight extension (or refinement) of the natural logic that evolution developed in human brains. Some postmodernists contest this claim, but their arguments are unconvincing since they ignore the above observation.

5. Almost all applications of mathematics can be developed in ZFC + $\varepsilon$. Hence, from the point of view of the present paper, ZFC + $\varepsilon$ and the operation $A(S)$ define a sufficiently large class of set theories **K**. However, it would be unfair not to mention that in recent decades a very significant development speaking in favor of a larger class **K** took place. J. Steel and W. H. Woodin proved that *a very natural class of sets, called $L(\mathbf{R})$, is an adequate universe for mathematical analysis, which is free of spurious pathological objects such a non-measurable sets in Euclidean spaces*. But their proof requires new axioms beyond ZFC + $\varepsilon$ asserting the existence of certain very large cardinal numbers. To be specific, the existence of one uncountable Tychonoff cardinal, suffices (see [6]). In view of the close analogy between such cardinals and the least infinite cardinal $\omega$, almost all set theorists think that it is consistent to assume their existence. Still, unlike the axioms of ZFC + $\varepsilon$ and additional axioms generated by the operation $A(S)$, these very large uncountable cardinals do not belong to the innate human logic. However, they are suggestive extensions of that logic that yield new and beautiful theorems.

**3. Planck's model.** As recalled in Chapter X of [5], Planck has shown that in order to derive his black body radiation curve it suffices to assume that excited atoms emit light in the form of quantum wave packets of microscopic duration. Thus, in this result, it is not needed to assume that atoms absorb light in quantum jumps; they can absorb it in a continuous way remaining resting most of their time in superposed states, but able to emit photons in quick bursts the moment they reach saturated states. And the packets they emit may have macroscopic wave fronts spreading in all directions. Electromagnetic radiation does not propagate in the form of particles but solely in the form of waves (which is the opposite of the view of Feynman). Let us generalize Planck's view and say that all detectors of particles have microscopic absorbers *A*, which signal only when they are filled by the energy or the matter carried by the radiation. Each *A* absorbs and attenuates only the portion of the wave that meets *A*.

Originally the levels at which the *A*'s are filled are in random (superposed) states uniformly distributed between being empty or full. When an *A* is in resonance with the incident radiation and does not reflect it, it is filled at a rate proportional to the absolute value of the square of the complex valued amplitude of the wave at the location of *A*.

Of course, this view explains the interference patterns of the two-slit experiment without assuming the "wave-particle duality".

In some media the quantum packets *P* emitted by *A*'s are focused along thin lines and can be observed as "particle tracks" since they cause microscopic signals tracing such tracks. If a wave packet *with a macroscopically extended front* moves through such a medium, it may cause in it secondary emissions that are focused along such tracks. (Planck did not say it, but I imagine that, in order to explain the Millikan oil drop experiment in his theory, one has to assume that in some media the *A*'s exchange their content and some of them reach saturated states while others become empty.)

It is important that, Planck's view constitutes a theory or a model in the sense of Section **2**. Indeed, this model tells us that: *the beams of radiation consist of real waves; when those waves propagate with macroscopic fronts they do not contain any localized particles; and the interactions between these waves and microscopic absorbers yield the microscopic corpuscular events*. Particles do not need to exist when they are not observed; wave packets do not collapse; space-time provides real separations between events. Thus, at least for a certain class of phenomena, the mystery of the wave-particle duality of the standard view of QM is eliminated. *However, Planck's theory has not been generalized to explain the whole realm of applications of* QM. For example, as much as I know, it does not explain a wave function over a phase spaces of a system of two interacting particles.

As mentioned in the introduction, in view of its limitations, it is generally believed that his model is false. Unfortunately, several popular arguments purporting to prove this are incomplete or faulty. The more credible ones are based on difficult relatively recent experiments and they necessitate delicate distinctions between superposed or entangled and mixed states; the mixed states are perfectly consistent with Planck's model, the entangled states are not. In the next section I will define an experiment that should offer such a direct proof. It still requires some sophisticated technology but it does not require these delicate distinctions.

Our experiment will focus on the simplest manifestation of non-locality,

which arises when a half-transparent mirror splits a beam of light in two directions. According to the standard interpretation, whenever a particle detector notices the wave packet of a photon in one of the branches, a particle detector in the other branch will miss that packet! However, as well known, if we recombine these branches and send them toward a photographic plate **F**, as in classical interferometry, then the dots that they produce on **F** will accumulate to form the familiar pattern of interference fringes. Thus, whether we attach any credence to Planck's model or not, we cannot agree with Feynman that *there are no waves*. Since we have to add their complex-valued amplitudes in order to predict the pattern we are compelled to believe that *there are real waves in both branches of the packet*.

But, as explained earlier, Planck's theory claims more, namely the absence of particles in any quantum packet that has an extended wave front. Hence his theory should be easier to refute. If it were true a single quantum packet should be able to produce more than one dot on **F**. Specifically, since the absorbers $A$ in **F** act independently, according to Poisson's formula, if the expected number of dots is $\lambda$, the probability of $n$ dots is $e^{-\lambda}\lambda^n/n!$ (n = 0, 1, 2, …).

**4. The experiment.** As in the experiments of A. Aspect et al [1], our source of light will be certain calcium atoms, which emit wave packets consisting of pairs of photons, a blue and a green photon in singlet state, moving in opposite directions. In the present experiment the only fact that matters is that in each such packet both photons are emitted simultaneously. (In [1] the additional fact, that the pair of photons is in singlet state, played an essential role; see Remark **6** in the next section.)

Let the blue photons go through a blue filter toward a detector $D_0$ and the green photons through a green filter toward a half transparent mirror **S**, that splits its part of the wave packet in two directions toward two other detectors $D_1$ and $D_2$ (see Figure). Planck's theory tells us that the detectors have some microscopic absorbers $A_1, \ldots, A_n$ (say atoms) such that each $A_i$ absorbs a part of each packet of the radiation, at a rate proportional to its intensity at $A_i$. It is only the secondary emissions (or signals) from the $A_i$ that are observed, and each $A_i$ signals within a microscopic time interval of the moment at which it became saturated, that is, when it reached an excited eigenstate. We need to secure that the packets emitted by our source do not overlap in time too often. Of course, Planck's model tells us that the detectors $D_1$ and $D_2$ act independently and this claim is inconsistent with the standard (Copenhagen) interpretation of QM, which says that $D_1$ and $D_2$ cannot signal the same packet.

Let $p_i$ be the conditional probability that $D_i$ ($i$ = 1, 2) detects a wave packet that is also detected by $D_0$, and $p_3$ be the conditional probability that both $D_1$ and $D_2$ detect a packet detected by $D_0$. Of course the Copenhagen Interpretation tells us that $p_3 = 0$ but Planck's model tells us that $p_3 = p_1 p_2$.

Of course, wave packets are not directly observable. Hence, we must define some measurable counterparts of the values $p_i$. Given a positive number $\alpha$, let $p_i(\alpha)$ ($i$ = 1, 2) be the conditional probability that a detection by $D_0$ at any time $t_0$ is accompanied by a detection by $D_i$ at a time $t_i$ such that $|t_0 - t_i| \leq \alpha$, for $i$ = 1, 2. Let $p_3(\alpha)$ be the conditional probability that a detection by $D_0$ at a time $t_0$ is accompanied by two detections by $D_1$ and by $D_2$ at times $t_1$ and $t_2$ such that $|t_1 - t_0| \leq \alpha$ and $|t_2 - t_0| \leq \alpha$. Then, by counting such single and such double detections in the time windows length $2\alpha$ defined by the detections of $D_0$, we can evaluate $p_1(\alpha)$, $p_2(\alpha)$ and $p_3(\alpha)$ with any desired accuracy (see Figure).

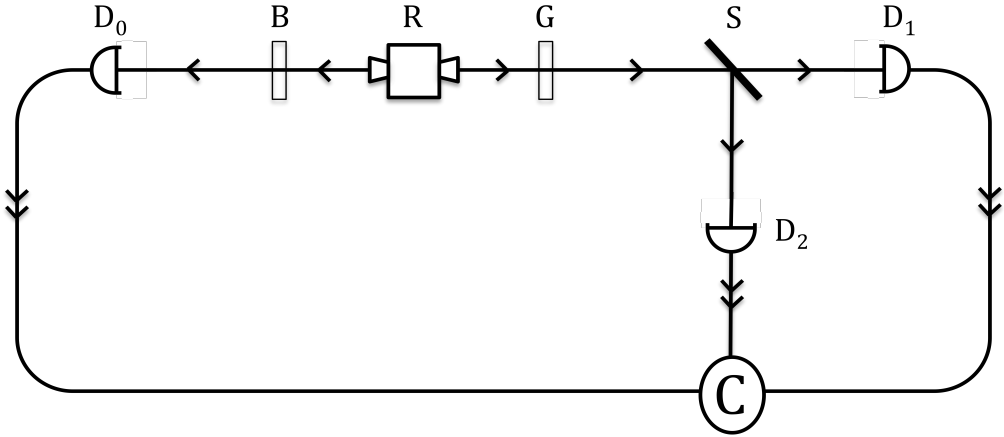

R source of light; ➤ paths of light; ➤➤ paths of signals;
B blue filter; G green filter; S half-transparent mirror;
$D_0$, $D_1$, $D_2$ photon detectors; C counter of signals estimating
the probabilities $p_1(\alpha)$, $p_2(\alpha)$, and $p_3(\alpha)$.

Figure

The Copenhagen view tells us that every wave packet caries only one green photon, thus, if the expected number of photons in the window $2\alpha$ is small enough, we will find that $p_3(\alpha)$ is close 0 or at least that $p_3(\alpha) < p_1(\alpha)p_2(\alpha)$. But Planck's model implies that $p_3(\alpha) = p_1(\alpha)p_2(\alpha)$ for all $\alpha$. Thus, if we can find $\alpha > 0$ such that $p_3(\alpha) < p_1(\alpha)p_2(\alpha)$ we will disprove his model and prove the entanglement of signals with absence of signals, i.e., an instantaneous communication between $D_1$ and $D_2$.

(Of course, the order of our color filters does not matter. We could exchange them sending the green photons toward $D_0$ and the blue photons toward $S$. But removing both filters would diminish our confidence that in each time window all the detectors $D_i$ signal only one entangled pair. Some variants of our experiment, that would use other technologies, and as much as I know could be easier to implement, are proposed below in Remark **4**.)

**5. Complementing remarks. 1.** As mentioned in Section **3** Planck's model is not refuted by the phenomenon of particle tracks. It suffices to accept that, quantum packets emitted in some media are focused along thin lines; these packets may be those of the original beam or they may be secondary emissions arising in the medium in which the tracks are observed. Also, in spite of what is often expressed in the literature, the photoelectric effects fail to prove the presence of particles, however they can demonstrate certain interactions between beams of radiation.

**2.** In our experiment it would be best if the packets were emitted by **R** do not overlap in time. But *this condition cannot be fully satisfied.* Thus let $p_0$ be the probability that, when **R** emits a quantum wave packet *P*, then it emits also other packets so close in time to *P* that, in our time window $2\alpha$, **D**$_1$ detects some of them and **D**$_2$ detects some others. However, it will suffice that $p_0$ be sufficiently small. Indeed, by the Copenhagen interpretation $p_3 \leq p_0$, while according to Planck's model $p_3 \geq p_1 p_2$ is still true. Hence our experiment will be significant if $p_0 < p_1 p_2$. In order to secure this inequality a source of light consisting of a single atom would be best and, of course, $\alpha$ should be small enough.

**3.** The blue branch and the detector **D**$_0$ are used only to mark such times at which a green packet goes to **D**$_1$ and **D**$_2$. Of course this is necessary for evaluating the conditional probabilities $p_1$ and $p_2$. Without this information the near absence of simultaneous detections by **D**$_1$ and **D**$_2$ could indicate merely that there are many packets *P* reaching them but each *P* is of very low intensity. Then the experiment could not distinguish the Copenhagen interpretation from Planck's model since both would say that the product $p_1 p_2$ is extremely small.

**4.** A possible variant of our experiment is the following. As well known beams of electrons of very low intensity, can accumulate dots on a photographic plate *P* in the form of macroscopic circular interference patterns. Hence in such beams individual packets are also dispersed over macroscopic wave fronts. Now, replace *P* by two independent electron detectors **D**$_1$ and **D**$_2$. Perhaps the absence or the presence of simultaneous detections of single electron packets by these **D**$_i$'s can be easier to establish than, as in the former experiment, for single photon packets.

Perhaps another variant would be also doable. Use quantum packets of single atoms or single molecules in some atomic or molecular beams, or some other heavy particles. Put a detector **D**$_0$ on their path between the source **R** and the beam splitting device **S**, such that **D**$_0$ signals their passage but lets them through toward **S**.

**5.** Modifying our experiment, we could study the shape of wave packets. Suffices to remove **S**, such that all the green photons go only to **D**$_1$. Then we can evaluate a conditional probability $p(s)$ that a photon detected by **D**$_0$ at time $t_0$ is accompanied by a detection by **D**$_1$ at a time $t_1$ such that $|t_0 + s - t_i| \leq \alpha$. The dependence of $p(s)$ on $s$ will inform us about the shape of the packets.

**6.** As mentioned earlier, the experiments of Aspect et al [1] were intended to prove a different phenomenon, namely, the existence of entangled pairs of photons. In mathematical terms their result can be expressed as follows. *There exist three Bernoulli two-valued random variables $X_1$, $X_2$, $X_3$, with $P(X_i = 0) = P(X_i = 1) = 1/2$ for i = 1, 2, 3, such that for each pair $i \neq j$ both $X_i$ and $X_j$ are simultaneously measurable and $P(X_i = X_j) = 1/4$.* This is very surprising since it implies that, in spite of simultaneous measurability of pairs, the three variables $X_i$ have no common distribution function, i.e., the standard Kolmogorov model does not exist for this triple. To prove this we need to show first that, if the whole triple had a common distribution function, then we would have $P(X_1 \neq X_0 \ \& \ X_2 \neq X_0) \geq 1/2$. Indeed, if we have two events such that each has the probability 3/4, then the probability of their conjunction must be $\geq 1/2$. Then, since $X_1 \neq X_0 \ \& \ X_2 \neq X_0$ implies $X_1 = X_2$, it follows that $P(X_1 = X_2) \geq 1/2$. This contradicts the former claim that $P(X_1 = X_2) = 1/4$.

**7.** As mentioned above the idea of our experiment is similar to these of [1] and [2], but the latter were designed to corroborate the existence of pairs of particles in entangled states. Our conditional probabilities $p_1(\alpha)$, $p_2(\alpha)$ and $p_3(\alpha)$ should be easier to evaluate quite precisely than the quantities measured in [1] and [2]. To refute Planck's view by means of the latter it is necessary to distinguish statistically the mixed states of pairs of distant photons, which are predicted by Planck's model, from the entangled states of such pairs, which are absent in his model. In our experiment we avoid this task.